\documentclass[10pt,letterpaper]{article}
\usepackage[top=0.85in,footskip=0.75in,marginparwidth=2in]{geometry}

\usepackage[utf8]{inputenc}

\usepackage{cite}

\usepackage{nameref}
\usepackage[hidelinks]{hyperref}

\usepackage[right]{lineno}

\usepackage{microtype}
\DisableLigatures[f]{encoding = *, family = * }

\raggedright
\usepackage{changepage}

\usepackage[aboveskip=1pt,labelfont=bf,labelsep=period,singlelinecheck=off]{caption}

\makeatletter
\renewcommand{\@biblabel}[1]{\quad#1.}
\makeatother

\usepackage{lastpage,fancyhdr,graphicx}
\usepackage{epstopdf}
\fancyheadoffset[L]{2.25in}
\fancyfootoffset[L]{2.25in}

\usepackage{color}

\definecolor{Gray}{gray}{.25}

\usepackage{graphicx}

\usepackage{sidecap}

\usepackage{wrapfig}
\usepackage[pscoord]{eso-pic}
\usepackage[fulladjust]{marginnote}
\reversemarginpar

\begin{document}
\vspace*{0.35in}

{\Large
\textbf\newline{MOSGA: Modular Open-Source Genome Annotator}
}
\newline
\\
Roman Martin\textsuperscript{1,2},
Thomas Hackl\textsuperscript{3},
George Hattab\textsuperscript{1},
Matthias G. Fischer\textsuperscript{3},
Dominik Heider\textsuperscript{1,*}
\\
\bigskip
\bf{1} Department of Mathematics and Computer Science, University of Marburg, 35032 Marburg, Germany
\\
\bf{2} Department of Organic-Analytical Chemistry, TUM Campus Straubing, 94315 Straubing, Germany
\\
\bf{3} Max Planck Institute for Medical Research, Department of Biomolecular Mechanisms, 69120, Heidelberg, Germany
\bigskip
* dominik.heider@uni-marburg.de

\section*{Abstract}
The generation of high-quality assemblies, even for large eukaryotic genomes, has become a routine task for many biologists thanks to recent advances in sequencing technologies. However, the annotation of these assemblies - a crucial step towards unlocking the biology of the organism of interest - has remained a complex challenge that often requires advanced bioinformatics expertise. \\
Here we present MOSGA, a genome annotation framework for eukaryotic genomes with a user-friendly web-interface that generates and integrates annotations from various tools. The aggregated results can be analyzed with a fully integrated genome browser and are provided in a format ready for submission to NCBI. MOSGA is built on a portable, customizable, and easily extendible Snakemake backend, and thus, can be tailored to a wide range of users and projects. \\
We provide MOSGA as a publicly free available web service at https://mosga.mathematik.uni-marburg.de and as a docker container at registry.gitlab.com/mosga/mosga:latest. Source code can be found at https://gitlab.com/mosga/mosga \\

\section*{Introduction}
Over the last twenty years, whole-genome sequencing and analysis has emerged as an essential and widely used technique across life sciences. In particular, the sequencing of new microbial genomes is now standard practise and has accelerated discoveries into microbial diversity and evolvability, providing new insight into microbiome function, human health, and ecology. 

The technical advances accelerating the generation of genome assemblies have also increased the need for their efficient annotation in terms of genes and other features. There are a few genome annotation pipelines available, e.g., PASA~\cite{Haas2003}, MAKER \cite{Holt2011}, and Funannotate~\cite{Love2020}. PASA and Funannotate were developed for plant and fungal genome annotations, respectively. In contrast, MAKER is universal and flexible in terms of modularity and extensibility. However, all them use command-line interfaces and lack a graphical user interface (GUI), limiting their usability to trained bioinformaticians. Furthermore, these pipelines use strict workflows with predefined tools and parameters, which cannot easily be tailored to non-model organisms such as eukaryotic protists~\cite{Sibbald2017}. 

To overcome these limitations, we have developed the Modular Open-Source Genome Annotator (MOSGA), which has recently used to annotate protists genomes~\cite{Hackl2020}. MOSGA enables the easy creation of draft eukaryotic genome annotations by providing a GUI with several task-specific prediction tools and a set of Snakemake workflow rules. As to our knowledge, MOSGA is the first modular, freely-available genome annotation framework and pipeline with a modular graphical user interface.

\section*{Software description}
The implementation of the MOSGA pipeline comprises three layers (see Figure~S1): (A) the graphical web-interface, (B) the Snakemake workflow engine~\cite{Koster2012}, and (C) the data accumulator. 
The web-interface allows pipeline submission, execution, and job order management. According to a JSON rule file, the set of tools, parameters, filter options, and supporting information are dynamically created at the interface. Extensions to the interface can be made by changing only the rules in the JSON file. 
The Snakemake pipeline will apply the corresponding job-dependent rules out of our set of 63 predefined rules (see Table S1). The Snakemake workflow engine ensures optimal use of computational resources and guarantees a successful pipeline execution. Before the actual run, an exact representation of the task-specific pipeline is generated as a graph by Snakemake. An example is shown in Figure S2. The MOSGA framework can be extended at this layer by defining additional rules for new tools, parameters, or even filters.
The data accumulator is responsible for reading every single output from every selected tool and finally writing the corresponding output. Internally, the accumulator stores information into highly abstracted objects retrieved from several classes that comprehensively read-in different outputs. It unifies, sorts, and filters all retrieved information and additionally performs quality checks. After each tool has been executed, the accumulator writes the final genome feature table and a SQN file, that can be used for NCBI GenBank submissions. Moreover, several workflow rules enable the integration of the prediction outputs into JBrowse for visualizing the annotation results~\cite{Buels2016}. New input or output formats can easily be implemented by providing new reader or writer classes. Moreover, pre-implemented python classes for reading in standard formats like CSV or GFF facilitates the development of extensions.
MOSGA is freely available and hosted at \href{https://mosga.mathematik.uni-marburg.de}{mosga.mathematik.uni-marburg.de}. In addition, we provide a Docker file to allow the local deployment of the whole framework.

\section*{Results}
The MOSGA framework includes state-of-the-art predictions tools for genome annotations that were previously used for annotating four draft genomes by \cite{Hackl2020}. To extend the applicability of MOSGA to other projects, we included additional tools as described here:

MOSGA uses WindowMasker and RepeatMasker for genome soft-masking and repeats detection \cite{Morgulis2006,Smit2013}. Moreover, we integrated four popular protein-coding gene prediction tools, namely Augustus, BRAKER, GlimmerHMM, and SNAP \cite{Stanke2005,Hoff2016,Korf2004,Majoros2004}. Furthermore, we provide a workflow-specific figure (Fig S3) to help users choose the the best-fitting tool for \textit{ab~initio} predictions tasks based on a benchmark recently performed by \cite{Scalzitti2020}. For the preparation of RNA-seq data, we also integrated TopHat2 and HiSat2 as alignment tools \cite{Kim2013,Kim2015}. An example of an RNA-seq based annotation is available online on the MOSGA site. Functional annotations can be carried out via EggNog~5 \cite{Huerta-Cepas2019} and Swiss-Prot \cite{Bairoch2004}, tRNAs are predicted by tRNAscan-SE 2 \cite{Lowe2016}, and ribomsomal RNA is identified via SILVA \cite{Quast2013}.

\section*{Discussion}
Unlike other genome annotation pipelines and frameworks, MOSGA's intuitive interface enable the user to choose suitable applications depending on the scientific question and task at hand. This permits users to build their own task-specific pipeline or workflow and addresses a wide community. Moreover, the modularity of the MOSGA framework allows quick extensions with new tools and modifications. MOSGA can be directly used to prepare NCBI-compliant submissions and the underlying Snakemake workflow engine guarantees full reproducibility. 

\section*{Acknowledgments}

This work was supported by the BMBF-funded de.NBI Cloud within the German Network for Bioinformatics Infrastructure (de.NBI) (031A537B, 031A533A, 031A538A, 031A533B, 031A535A, 031A537C, 031A534A, 031A532B). 
We thank Marius Welzel for his best practices advice to Snakemake workflows.

\section*{Funding}
This study was funded by the European Regional Development Fund, EFRE-Program, European Territorial Cooperation (ETZ) 2

\nolinenumbers
\clearpage

\bibliography{ms}
\newpage

\bibliographystyle{abbrv}

\end{document}